\documentclass[pra,twocolumn,superscriptaddress,showpacs,preprintnumbers,amsmath,amssymb,aps]{revtex4}
\usepackage{eepic,pstricks}
\usepackage{fancybox}
\usepackage[dvips]{graphicx}
\usepackage{amsmath,amssymb,mathrsfs}
\usepackage{epic,eepic}
\usepackage{epsfig}
\def\Tr{\operatorname{Tr}} \def\d{\operatorname{d}}
\def\>{\rangle}\def\<{\langle} 
   
\begin{document}

\title{Information-Disturbance Tradeoff in Quantum State
  Discrimination} 

\author{Francesco Buscemi}
\affiliation{ERATO-SORST Quantum Computation and Information Project,
Japan Science and Technology Agency, Tokyo, Japan}
\affiliation{Dipartimento di Fisica ``A. Volta'' and CNISM, via Bassi 6,
I-27100 Pavia, Italy.}
\author{Massimiliano F. Sacchi}
\affiliation{Dipartimento di Fisica ``A. Volta'' and CNISM, via Bassi 6,
I-27100 Pavia, Italy.}
\affiliation{CNR - Istituto Nazionale per la Fisica della Materia, 
Unit\`a di Pavia, Italy.}

\date{\today}% It is always \today, today,

\begin{abstract}
  When discriminating between two pure quantum states, there exists a
  quantitative tradeoff between the information retrieved by the
  measurement and the disturbance caused on the unknown state. We
  derive the optimal tradeoff and provide the corresponding quantum
  measurement. Such an optimal measurement smoothly interpolates
  between the two limiting cases of maximal information extraction and
  no measurement at all.
\end{abstract}

\maketitle

%%%%%%%%%%%%%%%%%%%%%%%%%%%%%%%%%%%%%%%%%%%%%%%%%%%%%%%%%%%%%%%%%%%%%%
\section{Introduction}
The problem of discriminating between two different quantum states
reveals two main features that make Quantum Theory so much different
from the classical intuition. First, quantum state discrimination
involves \emph{in-principle} indistinguishability of quantum states:
it is well known that it is not possible to perfectly infer (by means
of a one-shot experiment) which state we eventually picked at random
from a set of non orthogonal quantum states. Of course, it is
nonetheless possible to perform such a decision in an optimal way,
e.~g., by minimizing the error probability of
discrimination~\cite{helstrom}, by a minimax strategy where the
smallest of the probabilities of correct detection is
maximized~\cite{mmax}, or looking for optimal unambiguous
discrimination~\cite{unambiguous}, where unambiguity is paid by the
possibility of getting inconclusive results from the measurement.
Second, quantum indistinguishability principle is closely related to
another very popular---yet often misunderstood---principle (formerly
known as Heisenberg principle~\cite{heisenberg,fuco,banaszek01.prl}): it is not
possible to extract information from a quantum system without
perturbing it somehow. In fact, if the experimenter could gather
information about an unknown quantum state without disturbing it at
all, even if such information is partial, by performing further
non-disturbing measurements on the same system, he could finally
determine the state, in contradiction with the indistinguishability
principle~\cite{gdm-yuen}.

Actually, there exists a precise tradeoff between the amount of
information extracted from a quantum measurement and the amount of
disturbance caused on the system, analogous to Heisenberg relations
holding in the preparation procedure of a quantum state.  Quantitative
derivations of such a tradeoff have been obtained in the scenario of
quantum state estimation \cite{hol,qse}. The optimal tradeoff has been
derived in the following cases: in estimating a single copy of an
unknown pure state \cite{banaszek01.prl}, many copies of identically
prepared pure qubits \cite{banaszek01.pra} and qudits \cite{mf}, a
single copy of a pure state generated by independent phase-shifts
\cite{mista05.pra}, an unknown maximally entangled state \cite{max},
an unknown coherent state \cite{cv} and Gaussian state \cite{paris}.
Experiment realization of minimal disturbance measurements has been
also reported \cite{dema,cv}.

The present paper aims at fully characterize such a tradeoff relation
in quantum state discrimination, in the case in which the unknown
quantum state is chosen with equal \emph{a priori} probability from a
set of two non orthogonal pure states, and the error probability of
the discrimination is allowed to be suboptimal (thus intuitively
causing less disturbance with respect to the optimal
discrimination). We explicitly provide a measuring strategy---both in
terms of outcome probabilities and state-reduction---that achieves the
optimal tradeoff, which smoothly interpolates between the two limiting
cases of maximal information extraction and no measurement at all. As
a byproduct, we also recover in a simpler way some of the results of
Ref. \cite{fuco}. Our explicit derivation of the quantum measurement
should allow to carry out a feasibility study for the experimental
realization of minimal-disturbing measurements.  The issue of the
information-disturbance tradeoff for state discrimination can become
of practical relevance for posing general limits in information
eavesdropping and for analyzing security of quantum cryptographic
communications. 

The paper is organized as follows. In Sec. II, we briefly review the problem of
minimum-error state discrimination, and obtain the minimum disturbance
for the minimum-error measurement. In Sec. III, we provide the general
solution of the optimal information-disturbance tradeoff, along with
the corresponding measurement instrument. In Sec. IV, we suggest an
experimental realization of the minimum-disturbing measurement and
conclude the paper with closing remarks.

\section{Minimum disturbance for minimum-error state discrimination}

Typically, in quantum state discrimination we are given two (fixed)
non orthogonal pure states $\psi_1$ and $\psi_2$, with \emph{a priori}
probabilities $p_1$ and $p_2=1-p_1$, and we want to construct a
measurement discriminating between the two. In the following, in order
to work in full generality, we will describe a measurement by means of
the \emph{quantum instruments} formalism~\cite{davies-lewis}, namely,
a collection of completely positive maps $\{\mathcal{E}_i\}$, labelled
by the measurement outcomes $\{i\}$. By exploiting the well known
Kraus decomposition~\cite{kraus}, one can always write
$\mathcal{E}_i(\rho)=\sum_kE^{(i)}_k\rho E^{(i)\dag}_k$. In the case
the sum comprises just one term, namely, $\mathcal{E}_i(\rho)=E_i\rho
E^\dag_i$, the map $\mathcal{E}_i$ is called \emph{pure}, since it
maps pure states into pure states. The trace
$\Tr[\mathcal{E}_i(\rho)]=\Tr[\Pi_i\rho]$, where
$\Pi_i=\sum_kE^{(i)\dag}_kE^{(i)}_k$ is a positive operator associated
to the $i$-th outcome, provides the probability that the measurement
performed on a quantum system described by the density matrix $\rho$
gives the $i$-th outcome. The posterior (or reduced) state after the
measurement is given by
$\rho_i=\mathcal{E}_i(\rho)/\Tr[\mathcal{E}_i(\rho)]$. The averaged
reduced state---coming from ignoring the measurement outcome---is
simply obtained using the \emph{trace-preserving} map
$\mathcal{E}=\sum_i\mathcal{E}_i$. The trace-preservation constraint
for $\mathcal{E}$ implies that the set of positive operators
$\{\Pi_i\}$ is actually a positive operator-valued measure (POVM),
satisfying the completeness condition $\sum_i\Pi_i=\openone$.

Quantum state discrimination is then performed by a two-outcome
instrument $\{\mathcal{E}_1,\mathcal{E}_2\}$ whose capability of
discriminating between $\psi_1$ and $\psi_2$ can be evaluated by the
average success probability
\begin{equation}\label{eq:prob}
\begin{split}
  P(\{\mathcal{E}_1,\mathcal{E}_2\})&=\sum_{i=1}^2p_i
  \Tr[\mathcal{E}_i(|\psi_i\>\<\psi_i|)]\\
  &=\sum_{i=1}^2p_i\Tr[\Pi_i|\psi_i\>\<\psi_i|].
\end{split}
\end{equation}
Notice that $P$ actually depends only on the POVM $\{\Pi_i\}$. The
probability $P$ quantifies the amount of information that the
instrument $\{\mathcal{E}_1,\mathcal{E}_2\}$ is able to extract from
the ensemble $\{p_1,\psi_1;p_2,\psi_2\}$. Among all instruments
achieving average success probability $\bar P$ (the bar over $P$ means
that we fix the value of $P$), we are interested in those minimizing
the average disturbance caused on the unknown state, that we evaluate
in terms of average fidelity, namely,
\begin{equation}\label{eq:disturb}
  D(\{\mathcal{E}_1,\mathcal{E}_2\},\bar P)=1-\sum_{i=1}^2p_i\<\psi_i|\mathcal{E}(|\psi_i\>\<\psi_i|)|\psi_i\>.
\end{equation}
Differently from $P$, the disturbance $D$ strongly depends on the
particular form of the instrument $\{\mathcal{E}_i\}$. This means that
there exist many different instruments achieving the same $P$, but
giving different values of $D$. Let
\begin{equation}\label{eq:hard-min}
  \bar D(\bar
  P)=\min_{\{\mathcal{E}_1,\mathcal{E}_2\}}
  D(\{\mathcal{E}_1,\mathcal{E}_2\},\bar P)
\end{equation}
be the disturbance produced by the \emph{least disturbing} instrument
that discriminates $\psi_1$ from $\psi_2$ with average success
probability $\bar P$. Intuitive arguments suggest that the larger is
$\bar P$, the larger must correspondingly be $\bar D$ (i.~e., the
larger is the amount of information extracted, the larger is the
disturbance caused by the measurement). Our aim is to quantitatively
derive such a tradeoff, along with the corresponding measurement instrument. 
From now on we will restrict to the case of
equal \emph{a priori} probabilities, i.~e. $p_1=p_2=1/2$.

\begin{figure}[htb]
\epsfig{file=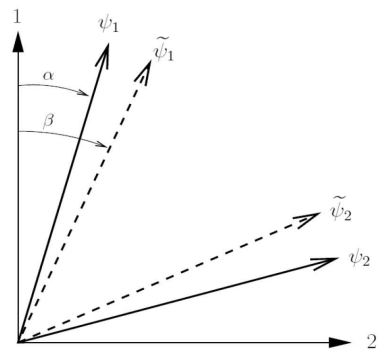,width=5cm}
\caption{Helstrom's scheme to optimally discriminate between to non
  orthogonal states $\psi_1$ and $\psi_2$. The orthogonal axes $1$ and
  $2$ correspond to the von Neumann measurement that 
  achieves the optimal discrimination probability
  (\ref{eq:helstrom}). According to the measurement outcome, 
  $\widetilde\psi_1$ and $\widetilde\psi_2$ are the states to
  be prepared, in order to
  minimize the disturbance, see Eq.~(\ref{eq:tilt}).}
\label{fig:quadrante}
\end{figure}

Let us start reviewing the case of the measurement maximizing $P$.
First of all notice that, given two generally non orthogonal pure
states $\psi_1$ and $\psi_2$, it is always possible to choose an
orthonormal basis $\{|1\>,|2\>\}$, placed symmetrically around
$\psi_1$ and $\psi_2$ (see Fig.~\ref{fig:quadrante}), on which both
states have real components, namely
\begin{equation}
\begin{split}
  &|\psi_1\>=\cos\alpha\;|1\>+\sin\alpha\;|2\>,\\
  &|\psi_2\>=\sin\alpha\;|1\>+\cos\alpha\;|2\>,
\end{split}
\end{equation}
and fidelity $f =|\<\psi_1|\psi_2\>|=\sin 2\alpha$. In this case,
it is known~\cite{helstrom} that the maximum achievable $P$ is
\begin{equation}\label{eq:helstrom}
  P_\textrm{opt}=\cos^2\alpha  ,
\end{equation}
which is obtained by the orthogonal von Neumann measurement
$\{|1\>\<1|,|2\>\<2|\}$.

Which is the instrument, among all instruments achieving
$P_\textrm{opt}$, that minimizes the disturbance $D$? Let us assume
for the moment (the optimality of this assumption will be proved in
full generality in the second part of the paper) that such an
instrument is pure. Intuitively, this means that we are excluding a
classical shuffling of outcomes. Then, since $P_\textrm{opt}$ is
reached by a rank-one von Neumann measurement, we can write
\begin{equation}\label{eq:Ui}
  \mathcal{E}_i(\rho)=U_i|i\>\<i|\rho|i\>\<i|U^\dag_i,\qquad i=1,2,
\end{equation}
where $U_i$ is a unitary operator. Letting
$U_i|i\>=|\widetilde\psi_i\>$, one recognizes in Eq.~(\ref{eq:Ui}) a
measure-and-prepare realization: the observable $|i\>\<i|$ is measured
and, depending on the outcome, the quantum state $\widetilde\psi_i$ is
prepared, i.~e. one has $\mathcal{E}_i(\rho)
=|\widetilde\psi_i\>\<\widetilde\psi_i|\Tr[\rho|i\>\<i|]$.  By
symmetry arguments (under the label exchange ``$1$'' $\leftrightarrow$
``$2$''), $U_1=U_2^\dag$, namely, the $\widetilde\psi_i$'s are
symmetrically tilted with respect to the $\psi_i$'s, see
Fig.~\ref{fig:quadrante}. With this notation, $D$ can be rewritten as
\begin{equation}\label{eq:opt-dist-to-min}
  D(P_\textrm{opt})=1-
\frac 12\sum_{i,j=1}^2|\<j|\psi_i\>|^2|\<\psi_i|\widetilde\psi_j\>|^2.
\end{equation}
Since $|\<i|\psi_i\>|^2=\cos^2\alpha$,
$|\<\psi_i|\widetilde\psi_i\>|^2=\cos^2(\beta-\alpha)$, and, for
$i\neq j$, $|\<j|\psi_i\>|^2=\sin^2\alpha$, and
$|\<\psi_i|\widetilde\psi_j\>|^2=\sin^2(\alpha+\beta)$, minimizing the
disturbance~(\ref{eq:opt-dist-to-min}) resorts to minimizing the
following function $D(\beta)$ of the tilt $\beta$ 
\begin{equation}\label{eq:max-disturbance}
  D(\beta)=1-\cos^2\alpha\cos^2(\beta-\alpha)-\sin^2\alpha\sin^2(\alpha+\beta),
\end{equation}
where $\alpha$ is a parameter, fixed along with the input states. 
Solving the equation $\d D(\beta)/\d\beta=0$, it turns out that the
tilt $\beta$ minimizing the disturbance is related to the angle
$\alpha$ by
\begin{equation}\label{eq:tilt}
\tan2\beta=\frac{\tan2\alpha}{\cos2\alpha},
\end{equation}
in agreement with Ref.~\cite{fuco}. From the above equation,
$\beta\ge\alpha$.
The presence of the tilt $\beta$ can be geometrically explained
starting from the observation that, for non orthogonal states, minimum
error discrimination can never be error-free. In other words, even
using the optimal Helstrom's measurement, there is always a non zero
error probability, and, the closer the input states are to each other,
the smaller the success probability is. Hence it is reasonable that,
the closer the input states are, the less ``trustworthy'' the
measurement outcome is, and the average disturbance is minimized by
cautiously preparing a new state that actually is a coherent
superpositions of both hypotheses $\psi_1$ and $\psi_2$. Using
Eq.~(\ref{eq:tilt}), from Eq.~(\ref{eq:max-disturbance}) one obtains
the minimum disturbance for Helstrom's optimal measurement
\begin{equation}\label{eq:disturbance_a}
  D_\textrm{opt}=\frac{4-\sqrt{14+2\cos 8\alpha }}{8}.
\end{equation}
Notice that $D_\textrm{opt}$ reaches its maximum for $\alpha=\pi/8$,
namely, when $\psi_1$ and $\psi_2$ are ``unbiased'' with respect to
each other ($|\<\psi_1|\psi_2\>|^2=1/2$).

\section{The general solution}
We analysed the limiting case in which the information extraction is
maximized---i.~e. the average success probability is maximized. The
opposite limiting case is when we do not perform any measurement at
all, without disturbing the states. The main result of the paper is to
provide the optimal tradeoff for all intermediate situations, along
with the corresponding quantum instrument. In order to do this, it is
useful to exploit the Choi-Jamio\l{}kowski isomorphism~\cite{choi-jam}
between \emph{completely positive} maps $\mathcal{M}$ on states on
$\mathcal{H}$ and \emph{positive} operators $R_\mathcal{M}$ on
$\mathcal{H}\otimes\mathcal{H}$ \begin{equation}\label{eq:choi-jam}
  \mathcal{M}\longleftrightarrow R_\mathcal{M}=(\mathcal{M}\otimes
\mathcal{I})|\Omega\>\<\Omega|
\end{equation}
where $|\Omega\>=\sum_{k=1}^d|k\>\otimes|k\>$ is the (non normalized)
maximally entangled vector in the $d^2$-dimensional Hilbert space
$\mathcal{H}\otimes\mathcal{H}$ (in our case, $\mathcal H$ is
two-dimensional). The correspondence~(\ref{eq:choi-jam}) is
one-to-one, the inverse formula being
\begin{equation}\label{eq:reconstruction}
  \mathcal{M}(\rho)=\Tr_2[(\openone\otimes\rho^*)\ R_\mathcal{M}],
\end{equation}
where $\Tr_2$ denotes the trace over the second Hilbert space, and
$\rho^*$ is the complex conjugated of $\rho$, with respect to the
basis fixed by $|\Omega\>$ in Eq.~(\ref{eq:choi-jam}). In terms of
Choi-Jamio\l{}kowski operator, trace-preservation condition is given
by $\Tr_1[R_\mathcal{M}]=\openone$.

An instrument $\{\mathcal{E}_1,\mathcal{E}_2\}$ can then be put in
correspondence with a set of positive operators $\{R_1,R_2\}$.
Clearly, $0<\Tr_1[R_1]<\openone$ and $0<\Tr_1[R_2]<\openone$, while
$\Tr_1[R_1+R_2]=\openone$, since the total operator $R=R_1+R_2$
corresponds to the trace-preserving map
$\mathcal{E}=\mathcal{E}_1+\mathcal{E}_2$. The average success
probability~(\ref{eq:prob}) and the average
disturbance~(\ref{eq:disturb}) can be rewritten as
\begin{eqnarray}
&&  P=\frac 12\sum_{i=1}^2\Tr[(\openone\otimes|\psi_i\>\<\psi_i|^*)\ R_i],
 \\
&&  
D=1-\frac 12\sum_{i=1}^2\Tr[(|\psi_i\>\<\psi_i|\otimes|\psi_i\>\<\psi_i|^*)\ R],
\label{14}
\end{eqnarray}
respectively. (In the following, we will drop the star, since
$\psi_1$ and $\psi_2$ have real components over the basis
$\{|1\>,|2\>\}$.) Our strategy is to fix the average success
probability $1/2\le P_t\le P_\textrm{opt}$ by fixing the value of a
control parameter $t$, i.~e.
\begin{equation}\label{eq:prob1-t}
  P_t=tP_\textrm{opt}+\frac{1-t}{2} 
= t\cos^2 \alpha +\frac{1-t}{2},
\end{equation}
with $0\le t\le 1$, and then to search, among all possible
measurements achieving $P_t$, for the one minimizing the disturbance
$D(P_t)$. In the symmetric case, $p_1=p_2=1/2$, the minimization
problem can be strikingly simplified by exploiting the exchange
symmetry $|\psi_1\>=\sigma_x|\psi_2\>$, where
$\sigma_x=\begin{pmatrix}0 & 1\\ 1& 0\end{pmatrix}$, in the
$\{|1\>,|2\>\}$ basis. It is then simple to check that, given an
instrument $\{R_1,R_2\}$ achieving average success probability $P$ and
disturbance $D$, the instrument $\{R'_1,R'_2\}$ constructed as
\begin{equation}
R'_i=\frac 12(R_i+\sigma_x^{\otimes 2}R_j\sigma_x^{\otimes 2}),\qquad i\neq j,
\end{equation}
achieves the same values of $P$ and $D$ as well. Hence, without loss
of generality, we can restrict ourselves to instruments satisfying
$R_2=\sigma_x^{\otimes 2}R_1\sigma_x^{\otimes 2}$. Then, the average
disturbance (\ref{14}) can be rewritten as $D=1-\Tr[\Sigma\ R_1]$,
where $\Sigma=\sum_{i}|\psi_i\>\<\psi_i|^{\otimes 2}$, and the
optimization problem~(\ref{eq:hard-min}) over a two-outcome instrument
resorts to the following---much simpler---optimization over a
\emph{single} positive operator $R_1$
\begin{equation}
  D_t=\min_{\{\mathcal{E}_1,\mathcal{E}_2\}} D(P_t)=1-\max_{R_1}\Tr[\Sigma\ R_1],
\end{equation}
with the trace-preservation constraint $\Tr_1[R_1+\sigma_x^{\otimes
  2}R_1\sigma_x^{\otimes 2}]=\openone$, and the constraint of average
success probability equal to $P_t$, namely
$\Tr[(\openone\otimes|\psi_1\>\<\psi_1|)\ R_1]=P_t$. These constraints
can be recast as four linear conditions:
\begin{eqnarray}
  &R_1\ge 0,\quad&\Tr[R_1]=1,\\
  &\Tr[(\openone\otimes\sigma_x)\ R_1]=0,\quad&\Tr[(\openone\otimes\sigma_z)\ R_1]=\frac{2P_t-1}{\cos2\alpha}\nonumber,
\end{eqnarray}
where $\sigma_z=\begin{pmatrix}1 & 0\\0 & -1\end{pmatrix}$.  Basic linear
programming methods, along with the reconstruction
formula~(\ref{eq:reconstruction}), show that, for every value of
$P_t$, the minimum disturbance $D_t$ is achieved by the \emph{pure}
instrument $\mathcal{E}_i^{(t)}(\rho)=E_i^{(t)}\rho E_i^{(t)\dag }$, 
where
\begin{equation}\label{eq:kraus-t}
\begin{split}
  E_1^{(t)}&=U(t)\left(\frac{\sqrt{1-\gamma}}{2}\sigma_z+\frac{\sqrt{1+\gamma}}{2}\openone\right),\\
  E_2^{(t)}&=U^\dag(t)\left(-\frac{\sqrt{1-\gamma}}{2}\sigma_z+\frac{\sqrt{1+\gamma}}{2}\openone\right),
\end{split}
\end{equation}
with $\gamma=\sqrt{1-t^2}$.  The unitary operator $U(t)$ in the above
equation generalizes that in Eq.~(\ref{eq:Ui}) as follows
\begin{equation}\label{eq:unitary-t}
U(t)=\begin{pmatrix}
\cos\beta_t & \sin\beta_t\\
-\sin\beta_t & \cos\beta_t
\end{pmatrix},
\end{equation}
where \cite{notafuco}
\begin{equation}\label{eq:tilt-t}
  \tan2\beta_t=\frac{t\sin2\alpha}{\cos^22\alpha+\gamma\sin^22\alpha}.
\end{equation} 
It follows that every instrument that achieves average success
probability $P_t$ must cause \emph{at least} an average disturbance
\begin{equation}\label{eq:dist1-t}
\begin{split}
  D_t&=\frac 12\left(1-t\sin2\alpha\sin2\beta_t\right)\\
  &+\frac{\cos2\beta_t}4\left[\gamma(\cos4\alpha-1)-\cos4\alpha-1\right].
\end{split}
\end{equation}
It is also simple to check that
\begin{equation}
\Pi_i^{(t)}=E_i^{(t)\dag}E_i^{(t)}=t|i\>\<i|+\frac{1-t}2\openone,
\end{equation}
namely,  the POVM of the measurement is the \emph{convex
  mixture} of the optimal one $\{|1\>\<1|,|2\>\<2|\}$ and a 
completely random one. On the
contrary, the instrument operators (\ref{eq:kraus-t}) represents a
\emph{coherent superposition} of Helstrom's (see Eq.~(\ref{eq:Ui}))
and the identity map.

Just by varying the control parameter $t$, it is possible to smoothly
move between the limiting cases. For $t=0$, we obtain the identity
map, that is, the no-measurement case. For $t=1$, we obtain Helstrom's
instrument in Eq.~(\ref{eq:Ui}), thus proving that assuming pure
instruments is in fact \emph{the} optimal choice.  In particular,
Eq.~(\ref{eq:tilt-t}) provides the tilt given in Eq.~(\ref{eq:tilt}).
However, the crucial difference between Helstrom's limit ($t=1$) and
the intermediate cases is that, for $t<1$, the optimal instrument
\emph{cannot} be interpreted by means of a measure-and-prepare scheme,
and the unitaries $U(t)$ and $U^\dag (t)$ in Eq. (\ref{eq:kraus-t})
represent feedback rotations for outcomes $1$ and $2$.

\begin{figure}[htb]
\epsfig{file=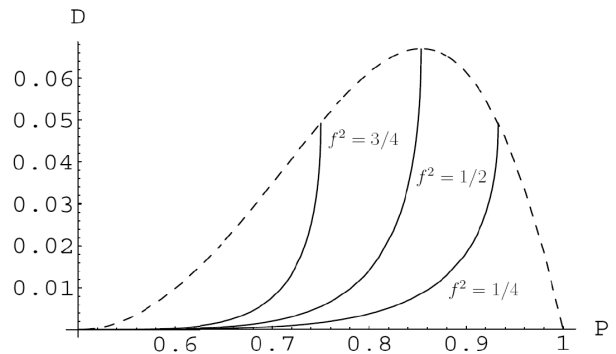,width=8cm}
\caption{Optimal tradeoff between disturbance and discrimination
  probability of two pure states $\psi _1$ and $\psi _2$ for three
  different values of the fidelity $f=|\<\psi_1|\psi_2\>|$, i.e.
  $f^2=\frac 34,\frac 12,\frac 14$. 
  The convolution of ending points (the dashed curve) provides the
  minimal disturbance for Helstrom's optimal measurement, namely
  $D_\textrm{opt}$ of Eq.~(\ref{eq:disturbance_a}), with $P=\cos^2
  \alpha$.}
\label{fig:tradeoff}
\end{figure}

By eliminating the parameter $t$ from Eqs.~(\ref{eq:prob1-t})
and~(\ref{eq:dist1-t}), we obtain the optimal tradeoff 
$D(P)$ between information and disturbance, for any value of $\alpha
$, namely for any couple of states with fidelity $f=\sin 2 \alpha
$. We plot $D(P)$ in Fig.~\ref{fig:tradeoff}, 
for three different values of $f$, i.e. $f^2=\frac 34, \frac 12, \frac
14$. 
 
The expression of $D(P)$ is rather involved, however it can be
simplified upon introducing the renormalized quantities 
$\mathscr{I}$ and $\mathscr{D}$ as follows 
\begin{equation}
  \mathscr I=\frac{P-1/2}{P_\textrm{opt}-1/2}, \qquad \mathscr
  D=\frac{D}{D_\textrm{opt}}, \end{equation} where $P_\textrm{opt}$
  and $D_\textrm{opt}$ are given in Eqs.  (\ref{eq:helstrom}) and
  (\ref{eq:disturbance_a}), respectively.  Clearly, one has
  $0\le\mathscr{I,D}\le 1$. After some lengthy algebra, we recover the
  following result of Ref. \cite{fuco} without any assumption: the
  optimal tradeoff between
  the amount of information $\mathscr I$ retrieved from the
  measurement and the disturbance $\mathscr D$ caused on the state is
  given by \begin{equation}
  \sqrt{D_\textrm{opt}\mathscr D(1-D_\textrm{opt}
    \mathscr D)}= 
\frac{\sin(4\alpha)}4\left(1-\sqrt{1-\mathscr I^2}\right).\label{28}
\end{equation}
For an optimal instrument, equality (\ref{28}) holds
for any value of $\alpha $, whereas for any suboptimal instrument the
l.h.s is strictly larger than the r.h.s. 

\section{Conclusion} In conclusion, a tight bound between the
probability of discriminating two pure quantum states and the degree
the initial state has to be changed by a quantum measurement has been
derived.  Such a bound can be achieved by a noisy measurement
instrument, where the noise continuously controls the tradeoff between
the information retrieved by the measurement and the disturbance on
the original state. More precisely, the optimal POVM is given by the
convex combination of the minimum-error POVM and the completely
uninformative one, whereas the measurement instrument is given by the
coherent superposition of the minimum-disturbing instrument for the
optimal discrimination and the identity map.

We finally suggest two possible experimental realizations of the
minimum-disturbing measurement, whose details will be published
elsewhere~\cite{else}. Since we are interested not only in the success
probability but also in the posterior state of the system \emph{after}
the measurement, we have to focus on a possible indirect measurement
scheme, in which the system is made interact with a probe, in such a
way they get entangled. After such interaction takes place, a
projective measurement is performed on the probe. The
\emph{mathematical} parameter $t$ controlling the tradeoff in
Eq.~(\ref{eq:prob1-t}) can then be put in correspondence with a
\emph{physical} parameter controlling the strength of the interaction
between the system and the probe: $t=0$ means that the interaction is
actually factorized in such a way that the following measurement on
the probe does not provide any information about the system and the
system is completely unaffected by the probe's measurement, that is,
the no-measurement case. On the contrary, $t=1$ identifies a
\emph{completely entangling} interaction, or, in other words, a
situation in which a measurement on the probe gives the largest amount
of information about the system, consequently causing the largest
disturbance. Two possible schemes for two-level systems encoded on
photons satisfy our requirements, that is, an entangling interaction
produced by means of a non-linear Kerr medium~\cite{ima}, or an
entangling measurement realized as a parity check~\cite{dema}. The
first approach, even if deterministic---i.~e.~no events have to be
discarded in principle---has serious drawbacks in reaching the value
$t=1$, since too large Kerr nonlinearity is needed \cite{ima}. 
On the other hand, the second approach is 
probabilistic---one half of the events are discarded---but it is based just on 
linear optics and it has been already implemented
and successfully tested~\cite{dema}. 

\section*{Acknowledgments} This work has been sponsored by Ministero
Italiano dell'Universit\`a e della Ricerca (MIUR) through FIRB (2001)
and PRIN 2005. F.~B. acknowledges Japan Science and Technology Agency
for partial support through the ERATO-SORST Project on Quantum
Computation and Information.

\appendix


\begin{thebibliography}{99}
\bibitem{notafuco}Equation (\ref{eq:tilt-t}) recovers a result of Ref.
  \cite{fuco}, where $\beta _t$ appears there in the expression of the
  disturbance. Here, it is shown where such a parameter comes from,
  namely it is explicitly related to the measurement instrument that
  we provide in Eqs. (\ref{eq:kraus-t})-(\ref{eq:unitary-t}).
\end{thebibliography}
\end{document}